------------------------------------------------------------------------
\documentstyle[11pt,preprint,aps]{revtex}

\begin{document}

\draft

\title{Coordinate-space Faddeev-Hahn-type  approach to
three-body charge transfer reactions involving exotic particles}

\author{ Renat A. Sultanov and Sadhan K. Adhikari}

\address{Instituto de F\'\i sica Te\'orica,
Universidade Estadual Paulista,\\
01405-900 S\~{a}o Paulo, S\~{a}o Paulo,
Brazil}

\maketitle

\vspace{2cm}

\begin{abstract}

Low-energy muon-transfer cross sections and rates in collisions of muonic
atoms with hydrogen isotopes are calculated using a six-state
close-coupling approximation to coordinate-space Faddeev-Hahn-type
equations.  In the muonic case satisfactory results are obtained for all
hydrogen isotopes and the experimentaly observed strong isotopic
dependence of transfer rates is also reproduced.  A comparison with
results of other theoretical and available experimental works is
presented.  The present model also leads to good transfer cross sections
in the well-understood problem of antihydrogen formation in
antiproton-positronium collision. 

 \end{abstract}

%
%

\narrowtext
\section{INTRODUCTION}
\label{sec:intro}

Charge-transfer reactions involving few particles in atomic physics are
very challenging and interesting from both theoretical and experimental
points of view and we study here the problem of charge transfer in some
atomic
reactions involving exotic particles.  Specifically, we study 
muon transfer in
D$-$H$_\mu$, T$-$H$_\mu$, and T$-$D$_\mu$ systems
where the suffix $\mu $
denotes a muonic atom with the electron replaced 
by a muon ($\mu^-$). We also
study the problem of antihydrohen ($\bar{\mbox H}$) formation
in antiproton-positronium collision with the
positronium (Ps) in an initial 1s state.

On the theoretical side, in these transfer reactions one needs to consider
rearrangement of a charged particle. Because of the Coulomb interaction
one needs a careful treatment of the dynamics for a correct description.
If one can identify the basic dynamical ingredients necessary for a
satisfactory description of these processes involving a small number of
particles, such a study will help us to formulate models in more complex
situations.  On the experimental side, the present study involving muon
and positron transfer is of current interest in the muon-catalyzed fusion
cycle 
\cite{Tresch2-1998,Tresch1-1998,Thalmann-1998,Jacot-1998,Mulhauser-1993,Breunlich-89}
and in the formation and study of the antihydrogen atom
\cite{Holz-1997,Charlton-94}.

Although there are some experimental measurements and several theoretical
investigations on these processes, there still remain discrepancies among
various studies.  Here we use a different theoretical approach based on a
detailed few-body dynamical consideration for a careful reinvestigation of
these three-body charge-transfer reactions. Traditionaly, such problems are
investigated by a tractable approximation scheme in the Schr\"odinger
framework, without explicitly considering a few-body dynamical equation. 
In addition to variational calculations, these schemes include
close-coupling, hyperspherical, and adiabatic approximations. Here we
would like to point out that the processes of muonic transfer reactions
and antihydrogen formation are three-body Coulombic rearrangement
collisions. Consequently, it seems reasonable that in addition to
approximations based on the Schr\"odinger equation, a detailed few-body
consideration is useful.
In what follows we develop a method, which is
based on detailed few-body equations rather than the effective potential
treatment employed in alternative investigations.

For the three-charged particle system, say TD$\mu$, only two
asymptotic configurations are possible, i.e. (D$\mu$)\ T and
(T$\mu$)\ D.  This suggests to write down a set of two
coupled
equations for components $\Psi_1$ and $\Psi_2$ of the wave function $\Psi
= \Psi_1 + \Psi_2$ \cite{Hahn-1968,Hahn-1972}
with each component carrying the
asymptotic boundary condition for a specific configuration. One such
equation with two components for the three-particle system was first
written by Hahn \cite{Hahn-1968} following the most general decomposition
of the three-body wave function into three components suggested by Faddeev
\cite{Faddeev-1961} and is usually referred to as the Faddeev-Hahn
equation \cite{Sultanov-1999}.  We solve the integro-differential form of
this equation by a six-state close-coupling approximation scheme which
consists in expanding the wave function components $\Psi_1$ and $\Psi_2$
in terms of eigenfunctions of subsystem Hamiltonians in initial and final
channels, respectively. The resultant coupled equation is then projected
on the expansion functions. After a partial-wave projection this leads to
a set of one-dimensional coupled equations for the expansion coefficients,
which is solved numerically.

Recently, there have been considerable theoretical and experimental
interests in the study of the muon-transfer reactions between hydrogen
isotopes in the muon catalyzed fusion cycle
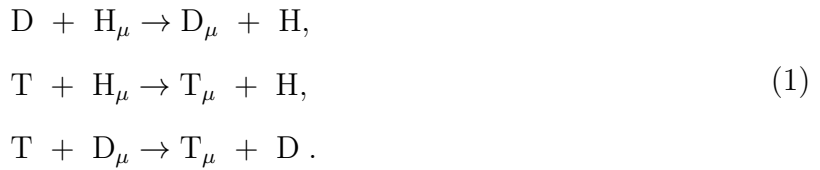
\begin{eqnarray}
\begin{array}{l}
\mbox{D} \ + \ \mbox{H}_{\mu} \rightarrow
\mbox{D}_{\mu} \ + \ \mbox{H}, \\
\mbox{T} \ + \ \mbox{H}_{\mu} \rightarrow
\mbox{T}_{\mu} \ + \ \mbox{H}, \\
\mbox{T} \ + \ \mbox{D}_{\mu} \rightarrow
\mbox{T}_{\mu} \ + \ \mbox{D}\;.
\label{eq:proc2}
\end{array}
\end{eqnarray}
The measurements for the transfer rates
\begin{equation}
\lambda_{\mbox{tr}} = \sigma_{\mbox{tr}} v N_0,
\label{eq:lam}
\end{equation}
with $\sigma_{\mbox{tr}}$ being the transfer cross section,
$v$ the relative velocity of the incident particles and
$N_0 = 4.25\times10^{22}$cm$^{-3}$ the liquid hydrogen density, are
listed in Table I together with recent
theoretical calculations. One can
see differences between different 
experimental data 
\cite{Dzhel-1962,Bleser-1963,Bertin-1972,Mulhauser-1996,Bystr-1981,Jones-83,Breun-87}
and theoretical results 
\cite{Cohen-1991,Adam-1992,Fukuda-1990,Kamimura-1993,Igarashi-1994,Lec-1997}.

One of the most attractive reactions for $\bar{\mbox{H}}$
formation is the three-body positron-transfer process
\begin{equation}
\bar{\mbox{p}} \ + \ \mbox{Ps} \rightarrow \bar{\mbox{H}} \ + \ e^-.
\label{eq:proc1}
\end{equation}
Although no experimental cross sections are available, this process is
being used at CERN for the production and study of antihydrogen.
A number of calculations
have recently been carried out to calculate
the cross section of reaction (\ref{eq:proc1}) as a function of the
incident Ps energy. The calculations were performed 
by different methods, for instance, with hyperspherical coupled-channel
expansions \cite{Toshima-1994} and close coupling approximations (CCA)
\cite{Mitroy-1997}.
As an additional test of the present method,
calculations for the S$-$wave
antihydrogen formation (\ref{eq:proc1}) at low energies are
also performed.

In Sec. II  we develop the formalism. The results
obtained for reactions (\ref{eq:proc2}) and (\ref{eq:proc1})
are given in
Sec. III. Finally, we present some  concluding remarks in Sec. IV.


\narrowtext
\section{Theoretical Formulation}

Let us take the system of units to be $e=\hbar=m_\mu=1$ and
denote, say T by ${\sf 1}$, D
by ${\sf 2}$ and muon by ${\sf 3}$.
Below the three-body breakup threshold, following 
two-cluster asymptotic configurations
are possible in the system ({\sf 123}):  $({\sf 23})\ -\ {\sf 1}$ and
$({\sf 13})\ -\ {\sf 2}$.
These configurations, denoted simply by 1 and 2, respectively,  are 
 determined by the Jacobi coordinates
$(\vec r_{j3}, \vec \rho_k)$
\begin{equation}
\vec r_{j3} = \vec r_3 - \vec r_j,
\hspace{6mm} \vec \rho_k =
(\vec r_3 + m_j\vec r_j) / (1 + m_j) - \vec r_k,
\hspace{6mm} j\not=k=1, 2,
\label{eq:coord}
\end{equation}
$\vec r_{j}$, $m_{j}$ are coordinates and
masses of the particles $j=1, 2, 3,$ respectively.

Let us introduce
the total three-body wave function as a sum of two components
\begin{equation}
\Psi(\vec r_1, \vec r_2, \vec r_3) \ =\  \Psi_1 (\vec r_{23},\vec \rho_1)
\ + \ \Psi_2 (\vec r_{13},\vec \rho_2),
\label{eq:total2}
\end{equation}
where $\Psi_1 (\vec r_{23},\vec \rho_1)$
is quadratically integrable over the variable
$\vec r_{23}$, and  $\Psi_2 (\vec r_{13},\vec \rho_2)$ over the
variable $\vec r_{13}$. 
To define $\Psi_l \, \, (l = 1, 2)$
a set of two coupled equations can be written down
\begin{eqnarray}
\begin{array}{l}
(E - H_0 - V_{23})\Psi_1 (\vec r_{23}, \vec \rho_1) =
(V_{23} + V_{12})\Psi_2 (\vec r_{13}, \vec \rho_2)
\vspace{7mm} \\
(E - H_0 - V_{13})\Psi_2 (\vec r_{13}, \vec \rho_2) =
(V_{13} + V_{12})\Psi_1 (\vec r_{23}, \vec \rho_1)\;,
\end{array}
\label{eq:fadd5}
\end{eqnarray}
where $E$ is the center-of-mass energy, $H_0$ is the total kinetic energy
operator,  and  $V_{ij} (r_{ij})$
are pair-interaction potentials $(i \not= j = 1, 2, 3)$.
Equations (\ref{eq:fadd5})  satisfy the Schr\"odinger
equation exactly and for  energies below the three-body
breakup threshold they possess the same advantages as the Faddeev
equations, since they are formulated for
the wave function components with correct physical
asymptotes.

In the general case a component of the three-body
wave function has the asymptotic form which includes all
open channels: elastic/inelastic, transfer and breakup.
In this case  each
component of the total wave function carries a specific  
asymtotic behavior.  The component $\Psi_1$ carries the asymptotic
behavior in elastic and inelastic channels:
\begin{equation}
\Psi_1(\vec r_{23}, \vec \rho_1)
\mathop{\mbox{\large$\sim$}}\limits_{\rho_1 \rightarrow + \infty}
e^{ik^{(1)}_1z}\varphi_1(\vec r_{23})\ + \ 
\sum_n A_n^{\mbox{\scriptsize{el/in}}}(\Omega_{\rho_1})
e^{ik^{(1)}_n\rho_1}/\rho_1\varphi_n(\vec r_{23})\; .
\end{equation}
The component $\Psi_2$ carries the asymptotic
behavior in the transfer channels:
\begin{equation}
\Psi_2(\vec r_{13}, \vec \rho_{2})
\mathop{\mbox{\large$\sim$}}\limits_{\rho_2 \rightarrow + \infty}
\sum_m A_m^{\mbox{\scriptsize{tr}}}(\Omega_{\rho_2})
e^{ik^{(2)}_m\rho_2}/\rho_2\varphi_m(\vec r_{13}),
\label{eq:tr}
\end{equation}
where $e^{ik^{(1)}_1z} \varphi_1(\vec r_{23})$ is the incident wave,
$\varphi_n(\vec r_{j3})$ the $n$-th excited bound-state wave function
of pair $(j3)$, $k_n^{(i)} = \sqrt{2M_i{(E- E_n^{(j)})}}$
with  $M_i^{-1}= m_i^{-1} + (1 + m_j)^{-1}$.
Here $E_n^{(j)}$
is the binding energy of  $(j3)$, $i\ne j = 1, 2$,
$A^{\mbox{\scriptsize{el/in}}}(\Omega_{\rho_{1}})$ and
$A^{\mbox{\scriptsize{tr}}}(\Omega_{\rho_{2}})$
are the scattering amplitudes in  the elastic/inelastic and transfer
channels.  This approach simplifies  the
solution procedure and simultaneously  provide a correct asymptotic
behaviour of the solution below the 3-body breakup threshold.

Let us write down Eqs. (\ref{eq:fadd5})
in terms of the adopted notations
\begin{eqnarray}
\left[ E + \frac{\nabla^2_{\vec \rho_k}}{2 M_k} +
\frac{\nabla^2_{\vec r_{j3}}}{2 \mu_{j}} -
V_{j3}\right] \Psi_k (\vec r_{j3}, \vec \rho_k)\ = \ 
%
(V_{j3} + V_{jk}) \Psi_{j}(\vec r_{k3}, \vec \rho_j)\;,
\label{eq:fadd8}
\end{eqnarray}
here $j\ne k =1, 2$ and
$M_k^{-1}= m_k^{-1} + (1 + m_j)^{-1}\ ,
\ \mu_j^{-1} = 1 + m_j^{-1}.$

For solving Eq. (\ref{eq:fadd8}) we expand the wave function
components in terms of bound states in initial and final channels,
and project this equation on these bound states. This prescription
is similar to that adopted in the close-coupling approximation.
Specifically, we use the following partial-wave expansion
\begin{equation}
\Psi_k(\vec r_{j3}, \vec \rho_{k}) = \sum_{LM\lambda l}
\Phi_{LM\lambda l}^{(k)}
(\rho_k, r_{j3}) \left \{ Y_{\lambda}(\hat \rho_k) \otimes
Y_l(\hat r_{j3}) \right \}_{LM}\; ,
\label{eq:lexpan}
\end{equation}
\begin{equation}
\left \{ Y_{\lambda}(\hat \rho_k) \otimes Y_l(\hat r_{j3}) \right \}_{LM}
=
\sum_{m'm}C^{LM}_{\lambda m'lm}
Y_{\lambda m'}(\hat \rho_k)Y_{lm}(\hat r_{j3}),
\end{equation}
where $C$'s are the Clebsch-Gordon coefficients, and $Y$'s are the usual 
spherical harmonics, and $L, \lambda, l$ and $M, m', m$ are the
appropriate angular momenta variables and their projections. Next we make
the
following close-coupling-type approximation for the radial part in terms
of the bound-state wave functions in the initial and final channels:
%
\begin{equation}
\Phi_{LM\lambda l}^{(k)}(\rho_k, r_{j3}) \approx
\frac{1}{\rho_k} \sum_n
f_{nl\lambda}^{(k)LM}(\rho_k) R_{nl}^{(k)}(r_{j3})\;,
\label{eq:expan}
\end{equation}
where radial components of the bound-state wave functions  
$R_{nl}^{(k)}(r_{j3})$ satisfy
\begin{equation}
\left \{E_n^{(k)} + \frac{1}{2\mu_j r_{j3}^2}
\left[\frac{\partial}{\partial r_{j3}} ( r_{j3}^2 \frac{\partial}
{\partial r_{j3}} ) - l(l+1)\right] - V_{j3} \right \}
R_{nl}^{(k)}(r_{j3})
= 0\;.
\label{eq:eqrnl}
\end{equation}

Then we substitute Eqs. (\ref{eq:lexpan})-(\ref{eq:expan})
into Eq. (\ref{eq:fadd8}), multiply the resultant equation  by the 
appropriate biharmonic functions and
the corresponding radial functions $R_{nl}^{(k)}(r_{j3})$, and
integrate over the corresponding angular coordinates of the vectors
$\vec r_{j3}$ and $\vec \rho_k$.
Then we obtain a set of integral differential equations for the
unknown functions $f_{nl\lambda }^{(k)}(\rho_k)$
\begin{eqnarray}
2M_k(E\ -\ E_n^{(j)})f_{\alpha}^{(k)}(\rho_k)\ +\ \left
\{\frac{\partial^2}
{\partial \rho_k^2}\ -\ 
\frac{\lambda (\lambda + 1)}{\rho_k^2}\right \} f_{\alpha}^{(k)}(\rho_k)
\ = \ 
2M_k \sum_{\alpha'}
\int_{0}^{\infty} dr_{j3} r_{j3}^2
\; \nonumber \\
\int d\hat r_{j3}
\int d\hat \rho_k
\frac{\rho_k}{\rho_j}
R_{nl}^{(k)}(r_{j3})
\left \{ Y_{\lambda}(\hat \rho_k) \otimes Y_l(\hat r_{j3}) \right \}_{LM}^*
(V_{j3} + V_{jk})
\left \{ Y_{\lambda '}(\hat \rho_j) \otimes Y_{l '}(\hat r_{k3})
\right \}_{LM}
\; \nonumber \\
R_{n'l'}^{(j)}(r_{k3})
f_{\alpha'}^{(j)}(\rho_j)\;.
\label{eq:part2}
\end{eqnarray}
For brevity we have  defined $\alpha \equiv nl\lambda$
and $\alpha^\prime \equiv n^\prime l^\prime \lambda^\prime$,
and omit the conserved total angular momentum label $LM$.
The functions $f_{\alpha}^{(k)}(\rho_k)$ 
depend on the scalar argument, but  Eq. (\ref{eq:part2}) is not yet
one-dimensional.
We are using the Jacobi coordinates 
\begin{eqnarray}
\vec \rho_j = \vec r_{j3} - \beta_k \vec r_{k3}, \hspace{4mm}
\vec r_{j3} = \frac{1}{\gamma} (\beta_k\vec \rho_k + \vec \rho_j),
\hspace{4mm}
\vec r_{jk} = \frac{1}{\gamma} (\sigma_j\vec \rho_j - \sigma_k
\vec \rho_k)\;,
\end{eqnarray}
with 
\begin{equation}
\beta_k = \frac{m_k}{1 + m_k},\hspace{4mm}\sigma_k = 1 - \beta_k,
\hspace{4mm} \gamma = 1 - \beta_k \beta_j,\hspace{4mm} j\not=k=1, 2. 
\end{equation}
This shows that modulus of $\vec \rho_j$ depend on two vectors:
$\vec \rho_j = \gamma \vec r_{j3} - \beta_k \vec \rho_k$. 
The  integration in the right-hand side   of Eq. (\ref{eq:part2}) 
is done over these two vectors.

To obtain one-dimensional integral differential equations,
corresponding to Eq.  (\ref{eq:part2}),
we proceed with the integration over variables
$\{\vec \rho_j,\vec
\rho_k\}$, rather than $\{\vec r_{j3},\vec \rho_k\}$. The Jacobian
of this transformation is $\gamma^{-3}$.
Thus, we come to a set of one-dimensional integral
differential equations
\begin{eqnarray}
2M_k(E - E_n^{(j)})f_{\alpha}^{(k)}(\rho_k) + \left \{\frac{\partial^2}
{\partial \rho_k^2} -
\frac{\lambda (\lambda + 1)}{\rho_k^2}
\right\} f_{\alpha}^{(k)}(\rho_k) = \; \nonumber \\
\frac{M_k}{\gamma^{3}}
\sum_{\alpha'}
\int_{0}^{\infty} d \rho_j
S_{\alpha \alpha'}^{(kj)}(\rho_k, \rho_j)
f_{\alpha'}^{(j)}(\rho_j)\;,
\label{eq:part4}
\end{eqnarray}
where functions $S_{\alpha \alpha'}^{(kj)}(\rho_k, \rho_j)$
are defined as follows
\begin{eqnarray}
S_{\alpha \alpha'}^{(kj)}(\rho_k, \rho_j) = 2\rho_k \rho_j
\int d \hat \rho_j \int d \hat \rho_k R_{nl}^{(k)}(r_{j3})
\left \{ Y_{\lambda}(\hat \rho_k) \otimes Y_l(\hat r_{j3}) \right \}_{LM}^*
(V_{j3} + V_{jk}) \nonumber \\
\times
\left \{ Y_{\lambda^\prime}(\hat \rho_j) \otimes Y_{l'}(\hat r_{k3})
\right \}_{LM} R_{n'l'}^{(j)} (r_{k3})\;.
\label{eq:ss}
\end{eqnarray}

The fourfold multiple integration in equations (\ref{eq:ss}) leads to
a singlefold integral and the expression (\ref{eq:ss})
for any value orbital momentum $L$ becomes 
\begin{eqnarray}
S_{\alpha \alpha'}^{(kj)}(\rho_k, \rho_j) =
\frac{4 \pi}{2L+1}
[(2\lambda + 1)(2\lambda^{\prime} + 1)]^{ \frac{1}{2} }
\rho_k \rho_j \int_{0}^{\pi}
d \omega \sin\omega R_{nl}^{(k)}(r_{j3})
(V_{j3}(r_{j3}) + V_{jk}(r_{jk}))\; \nonumber \\
R_{n'l'}^{(j)}(r_{k3})
\sum_{mm'} D_{mm'}^L(0, \omega, 0)
C_{\lambda 0lm}^{Lm}
C_{\lambda' 0l'm'}^{Lm'} Y_{lm}(\nu_j, \pi) Y^*_{l'm'}(\nu_k, \pi)\;,
\label{eq:final}
\end{eqnarray}
where $D_{mm'}^L(0, \omega, 0)$ are Wigner functions,
$\omega$ is angle between $\vec \rho_j$ and $\vec \rho_k$,
$\nu_j$ between $\vec r_{k3}$ and $\vec \rho_j$,
$\nu_k$ between $\vec r_{j3}$ and $\vec \rho_k$.

Finally,
the  set of integro-differential equations
for the unknown functions $f^{(k)}_{nl\lambda}(\rho_k)$ can be written as 
%
%
%
\begin{eqnarray}
\left[ (k^{(i)}_n)^2\ +\ \frac{\partial^2}
{\partial \rho_i^2}\ -\ 
\frac{\lambda (\lambda + 1)}{\rho_i^2}
\right] f_{\alpha}^{(i)}(\rho_i)\ =\ g_i \sum_{\alpha'}
\frac{\sqrt{(2\lambda + 1)(2\lambda^{\prime} + 1)}}{2L+1}
\; \nonumber \\
\int_{0}^{\infty} d \rho_{i'}
f_{\alpha^\prime}^{(i')}(\rho_{i'})\int_{0}^{\pi}
d \omega \sin\omega
R_{nl}^{(i)}(|\vec{r}_{i'3}|)
\left[-\frac{1}{|\vec {r}_{i'3}|} + \frac{1}{|\vec {r}_{ii'}|}\right]
R_{n'l'}^{(i')}(|\vec{r}_{i3}|)
\;\nonumber \\
\rho_{i} \rho_{i'} \sum_{mm'} D_{mm'}^L(0, \omega, 0)C_{\lambda 0lm}^{Lm}
C_{\lambda' 0l'm'}^{Lm'}
Y^{*}_{lm}(\nu_i, \pi) Y_{l'm'}(\nu_{i'}, \pi)\;.
\label{eq:most}
\end{eqnarray}
Here $i \not= i' = 1,2$,
$g_i =4\pi M_i/\gamma^{3}$,
$k^{(i)}_n = \sqrt{2M_i(E-E_n^{(i')})}$,
%
$\omega$ is angle between the Jacobi coordinates
$\vec \rho_i$ and $\vec \rho_{i'}$, $\nu_i$ is the angle between 
$\vec r_{i'3}$ and $\vec \rho_i$, $\nu_{i'}$ is angle
between $\vec r_{i3}$ and $\vec \rho_{i'}$ with
\begin{eqnarray}
\sin \nu_i = \frac{\rho_{i'}}{\gamma r_{i'3}}\sin\omega\; 
\hspace{10mm}
\mbox{and}
\hspace{10mm}
\cos \nu_i = \frac{1}{\gamma r_{i'3}}
(\beta_i \rho_{i} + \rho_{i'} \cos \omega).
\end{eqnarray}

To find unique solution to system (\ref{eq:most}),
appropriate boundary conditions are to be considered. First we impose
$f_{nl}^{(i)}(0) = 0$.
%
For the present scattering  problem with $1 +(23)$ as the initial state,
in the asymptotic region two solutions to Eq.(\ref{eq:most})
satisfy the following boundary conditions
\begin{eqnarray}
\left\{
\begin{array}{l}
f_{1s}^{(1)}(\rho_1)
\mathop{\mbox{\large$\sim$}}\limits_{\rho_1 \rightarrow + \infty}
\sin(k^{(1)}_1\rho_1) + {\it K}_{11}\cos(k^{(1)}_1\rho_1)\;,
\vspace{1mm}\\
f_{1s}^{(2)}(\rho_2)
\mathop{\mbox{\large$\sim$}}\limits_{\rho_2 \rightarrow + \infty}
\sqrt{v_1 / v_2}{\it K}_{12}\cos(k^{(2)}_1\rho_2)\;,\\
\end{array}\right.
\label{eq:cond88}
\end{eqnarray}
where 1 refer to channel ${\sf 1} + ({\sf 23})$, 2 to channel
${\sf 2} + ({\sf 13})$ and $K$ denotes the corresponding on-shell
${\bf K}$-matrix \cite{Mott-1965}. For scattering
with ${\sf 2} + ({\sf 13})$ as the initial state, we have the following
conditions
\begin{eqnarray}
\left\{
\begin{array}{l}
f_{1s}^{(1)}(\rho_1)
\mathop{\mbox{\large$\sim$}}\limits_{\rho_1 \rightarrow + \infty}
\sqrt{v_2 / v_1}{\it K}_{21}\cos(k^{(1)}_1\rho_1)\;,
\vspace{1mm}\\
f_{1s}^{(2)}(\rho_2)
\mathop{\mbox{\large$\sim$}}\limits_{\rho_2 \rightarrow + \infty}
\sin(k^{(2)}_1\rho_2) + {\it K}_{22}\cos(k^{(2)}_1\rho_2).
\end{array}\right.
\label{eq:cond8888}
\end{eqnarray}
where $v_i$, $i=1,2$ are velocities in channel $i$. 
With the following change of variables in Eqs.(\ref{eq:most})
\begin{equation}
{\sf f}_{1s}^{(1)}(\rho_1)=
f_{1s}^{(1)}(\rho_1)-\sin(k^{(1)}_{1}\rho_1)\hspace{6mm}
\mbox{and}\hspace{6mm}
{\sf f}_{1s}^{(2)}(\rho_2)=
f_{1s}^{(2)}(\rho_2)-\sin(k^{(2)}_{1}\rho_2),
\end{equation}
we can obtain two sets of inhomogeneous equations which are
solved numerically. The cross sections are given by
\begin{eqnarray}
\sigma_{ij} = \frac{4\pi}{k_1^{(i)2}}\left |\frac{{\bf K}}
{1 - i{\bf K}}\right |^2 = 
\frac{4\pi}{k_1^{(i)2}}\frac{\delta_{ij}D^2 + {\it K}_{ij}^2}
{(D - 1)^2 + ({\it K}_{11} + {\it K}_{22})^2},
\end{eqnarray}
where $i=j=1,2$ refer to the two channels and
\begin{equation}
D = \det {\bf K} = {\it K}_{11}{\it K}_{22} - {\it K}_{12}{\it K}_{21}.
\end{equation}
When $k_1^{(1)} \rightarrow 0$:
$K_{12} =  K_{21} \sim k_1^{(1)}$,
$K_{11} \sim k_1^{(1)}$, in this case
$\sigma_{\mbox{tr}} \equiv  \sigma_{12} \sim 1/k_1^{(1)},$
and $\sigma_{\mbox{el}} = \sigma_{11} \sim const.$
For comparison with experimental low-energy data it is very useful
to calculate the transfer rates (\ref{eq:lam}) because
$\lambda_{\mbox{tr}}(k_1^{(1)}\rightarrow 0) \sim const$.
%

\narrowtext
\section{Numerical Results}

To solve the integro-differential equation,  one has to calculate
the angle integrals in Eq. (\ref{eq:most}) which
are independent of the energy $E$.
One needs to calculate them only once and store on
hard disk for the  calculation of other observables, for instance, the
cross sections at different energies.
Subintegrals in Eq. (\ref{eq:most}) have strong dependence on
$\rho_i$ and $\rho_{i'}$ ($i \ne i'=1,2$). To calculate
$S_{\alpha \alpha'}^{(ii')}(\rho_i, \rho_{i'})$ at different coordinates
an adaptable algorithm has been used.  In this case using the
relation 
\begin{equation}
cos \omega = \frac{x^2 - \beta_i^2\rho_i^2 - \rho_{i'}^2}{2\beta_i
\rho_i\rho_{i'}}
\end{equation}
the angle dependent part of Eq. (20) can be written as the following
integral
\begin{eqnarray}
S_{\alpha \alpha'}^{(ii')}(\rho_i, \rho_{i'})= 
\frac{4\pi}{\beta_i}
\frac{[(2\lambda + 1)(2\lambda^{\prime} + 1)]^{\frac{1}{2}}}{2L+1}
\int_{|\beta_i\rho_i - \rho_{i'}|}^{\beta_i\rho_i + \rho_{i'}}
dx R_{nl}^{(i)}(x)
\left[-1 + \frac{x}{r_{ii'}(x)}\right]\;\nonumber \\
R_{n'l'}^{(i')}(r_{i3}(x))
\sum_{mm'} D_{mm'}^L(0, \omega(x), 0)C_{\lambda 0lm}^{Lm}
C_{\lambda' 0l'm'}^{Lm'}
Y^{*}_{lm}(\nu_i(x), \pi) Y_{l'm'}(\nu_{i'}(x), \pi).
\label{eq:omega}
\end{eqnarray}
Note that
the expression (\ref{eq:omega}) differs from zero only in a narrow
strip when $\rho_i \approx \rho_{i'}$.

We employ  
muonic atomic unit: distances are measured in units of $a_\mu$, where 
$a_\mu$ is the radius of muonic hydrogen atom.  
The integro-differential equations were solved by usual numerical
procedure by discretizing  them into a linear system of equations,
which are subsequently solved by Gauss elimination method. 
In solving these equations distances upto 50$a_\mu$ were considered
and 400 $-$ 600   points were used in the discretization. 
The following mass values are used in the unit of electron mass:
$m_H$ = 1836.152, $m_D$ = 3670.481, $m_T$ = 5496.918
and the muon mass is $m_{\mu}$ = 206.769.

Tables II, III, and IV  include our results for the muonic transfer
cross sections and rates
for all hydrogen isotopes (\ref{eq:proc2}) using different approximation
schemes. 
We present results for two-, four-, and six-state approximation where
we include 1s, 1s+2s and 1s+2s+2p states of the muonic atoms in the
initial and  final channels, respectively. In solving the equations we
employed only the lowest partial wave, e.g., $L=0$. As we shall mainly be
concerned with the experimental muon transfer rates at very low energies,
the higher partial waves are expected to have negligible contribution.  
The 2p states are found to contribute significantly in 
T-D$_\mu$, moderately in D-H$_\mu$, and little in T-H$_\mu$
systems. This is in agreement with similar  conclusion 
of Ref. \cite{Igarashi-1994}  in the T-D$_\mu$ system. 
This could be understood qualitatively from the following consideration.
At zero incident energy the relative velocity in the final state after
muon transfer is the highest in the case of T-H$_\mu$, lowest in the
case of T-D$_\mu$ and intermediate in the case of D-H$_\mu$. It is
expected that the polarization potential arising out of a 
1s+2s+2p calculation will have the largest effect on convergence when the
final-state velocity is the lowest. Hence the necessity of the
higher-order states 
is more pronounced in the case of T-D$_\mu$ and less pronounced in the
case of T-H$_\mu$. We also find that as energy decreases the transfer
cross sections increase and  the transfer rates attain a constant value. 
These  transfer rates are essentially constant below 0.1 eV and are also
measured experimentally, so that we can compare our rates with
other experimental  and  theoretical results. 

For the D-H$_\mu$
system the present low-energy muon transfer rate of 
133$\times 10^{8}$ s$^{-1}$ is in agreement with both experiments
\cite{Dzhel-1962,Bleser-1963}.
The present rate is slightly smaller than the theoretical studies  of
Refs. \cite{Cohen-1991},\cite{Adam-1992}
and this makes the agreement with experiment better.
For the T-H$_\mu$ system again
the present result 61$\times 10^{8}$ s$^{-1}$
is in better agreement with the
experiment \cite{Mulhauser-1996} than the other theoretical studies.
In case of T-D$_\mu$, the present result 2.3$\times 10^{8}$ s$^{-1}$
is also in very good agreement with experiment.

Within the six-state approximation our cross sections
for low energy elastic scattering in case  T-D$_\mu$ system
are presented in Table V together with other theoretical results.
The present
cross sections attain  a 
constant value at low energies and is in
fairly good
agreement with results of other studies.

As a futher test of the present few-body approach,
we have also calculated S-wave
cross sections of antihydrogen formation in antiproton-positronium
low energy collisions (\ref{eq:proc1}).  In Table VI our results
within the six state approximation
(Ps[1s+2s+2p],$\bar {\mbox H}$[1s+2s+2p])
are compared with calculations based on
hyperspherical coupled-channel method \cite{Toshima-1994}.
Considering that  the present calculation is limited to only the
lowest partial wave ($L=0$) and to a truncated basis set (1s+2s+2p),
the agreement is  reasonable for energies below   
1 eV. However, at 2 eV the agreement is not so good. The reason for
this is not clear at present.  Further theoretical investigation
including higher partial waves with an extended basis set could  
reveal the trend of the converged cross sections.


\narrowtext
\section{CONCLUSION}

The study of three-body Coulombic systems have been the subject of this
work.
We have formulated a method for a  few-body description of the
rearrangement scattering problem by solving the Faddeev-Hahn-type
equations in coordinate space. It is shown that within
this formalism, the application of a close-coupling-type ansatz
leads to satisfactory results already in low-order approximations
for (i) muon-transfer reactions between hydrogen isotopes and
(ii) antihydrogen 
formation in antiproton-positronium collision. 
Because of computation difficulties, in this preliminary
application we have
considered up to six states in the expansion scheme (1s+2s+2p on
each center), which may not always be adequate.
Further calculations with larger basis sets are
needed to obtain the converged  results.

The present model  leads to a
reduction of the usual technical effort and is definitely
worth using for investigations of larger systems. 
It seems reasonable to suppose that the method
should be an effective tool for the description of
other muonic and atomic few-body collisions.
For instance, one could study using the present approach the
following muon-transfer reactions to elements with $Z \ge 2$
\begin{equation}
(\mbox{H}_{\mu})_{1s} \ + \ \mbox{X}^Z
\rightarrow \mbox{X}^Z_{\mu} \ + \ \mbox{H},
\label{eq:proc7}
\end{equation}
where the cross section depends in a complicated manner on the charge
$Z$ \cite{Mulhauser-1993}.

Theoretically,
the reaction (\ref{eq:proc7}) is of much interest as an example
of low-energy rearrangement scattering in a system of three charged
particles with Coulomb repulsion in the final state. Evidently
it makes additional difficulties for correct theoretical
description of Eq. (\ref{eq:proc7}) \cite{Sultanov-1999}.
The Faddeev-Hahn-type
approach seems to be suitable for the study of such reactions and would be 
a topic of future investigation. We are presently in the process of 
studying reaction (\ref{eq:proc7}) with the present method for $Z
=2 $ and 3. We also plan to employ  an extended basis set with more 
basis functions in the future. Also, the
excited state muon-transfer reactions of recent experimantal
and theoretical interest  \cite{Lauss-96,Krav-1998}
could be studied with the present model.


\acknowledgments

We acknowledge the support from FAPESP (Funda\c{c}\~{a}o
de Amparo \~{a} Pesquisa do Estado de S\~{a}o Paulo) of  Brazil.
The numerical calculations have been performed on the IBM SP2
Supercomputer of the Departamento de
F\'\i sica - IBILCE - UNESP,
S\~{a}o Jos\'e do Rio Preto, Brazil.


%
%
\mediumtext
\begin{table}
\caption{Experimental and theoretical
results for the muonic transfer rates 
$\lambda_{\mbox{tr}}/10^{8}\mbox{s}^{-1}$
given for low energies ($E < 0.1$ eV);
*$-$ present results;
$^{\sharp} -$ rates reproduced from
cross sections.}
\vspace{3mm}
\begin{tabular}{lrrr}
  \multicolumn{1}{l}{Reaction}
& \multicolumn{1}{r}{Experiment}
& \multicolumn{2}{c}{Theory}\\
\tableline
D\ + \ H$_{\mu} \rightarrow $D$_{\mu}$ \ + \ H
  &95 $\pm$ 34 \cite{Dzhel-1962}&140 \cite{Cohen-1991}
  &159 \cite{Adam-1992}\\
  &143$\pm$ 13 \cite{Bleser-1963}& 133 [*]\\
  &84$\pm$  13 \cite{Bertin-1972}\\
\tableline
T \ + \ H$_{\mu} \rightarrow $T$_{\mu}$ \ + \ H
  &58.6$\pm$ 10 \cite{Mulhauser-1996}
  &55 \cite{Cohen-1991}
  &71.7 \cite{Adam-1992}\\
  & &61 [*]\\
\tableline
T \ + \ D$_{\mu} \rightarrow $T$_{\mu}$ \ + \ D
  &2.9$\pm$ 0.4 \cite{Bystr-1981}&3.5 \cite{Cohen-1991}
  &2.26 \cite{Adam-1992}\\  
  &2.8$\pm$ 0.3 \cite{Jones-83}&1.5$^{\sharp}$ \cite{Fukuda-1990}
  &2.8 \cite{Kamimura-1993}\\
  &2.8$\pm$ 0.5 \cite{Breun-87} &2.39$^{\sharp}$ \cite{Igarashi-1994}
  &0.93$^{\sharp}$ \cite{Lec-1997}\\  
  &3.5$\pm$ 0.5 \cite{Breun-87} &2.3 [*]&
\label{tab4}
\end{tabular}
\end{table}
\vspace{1cm}
%
%
%
\begin{table}
\caption{
Cross sections $\sigma(\mbox{D}$-$\mbox{H}_\mu) =
\sigma_{\mbox{tr}}/10^{-20}$ cm$^2$ and rates 
$\lambda(\mbox{D}$-$\mbox{H}_\mu) =
\lambda_{\mbox{tr}}/10^{10}\mbox{s}^{-1}$  for
$\mu$-transfer reaction
D \ + \ H$_{\mu} \rightarrow $D$_{\mu}$ \ + \ H, 
at different energies.}
\vspace{2mm}
\begin{tabular}{lccccccccc}
  \multicolumn{1}{l}{$E$ (eV)}
& \multicolumn{1}{c}{$\sigma(\mbox{D}$-$\mbox{H}_\mu)$}
& \multicolumn{1}{c}{$\lambda(\mbox{D}$-$\mbox{H}_\mu)$}
& \multicolumn{1}{c}{ }
& \multicolumn{1}{c}{$\sigma(\mbox{D}$-$\mbox{H}_\mu)$}
& \multicolumn{1}{c}{$\lambda(\mbox{D}$-$\mbox{H}_\mu)$}
& \multicolumn{1}{c}{ }
& \multicolumn{1}{c}{$\sigma(\mbox{D}$-$\mbox{H}_\mu)$}
& \multicolumn{1}{c}{$\lambda(\mbox{D}$-$\mbox{H}_\mu)$}\\
  \multicolumn{1}{l}{ }
& \multicolumn{2}{c}{1s}
& \multicolumn{1}{c}{ }
& \multicolumn{2}{c}{1s+2s}
& \multicolumn{1}{c}{ }
& \multicolumn{2}{c}{1s+2s+2p}\\
\hline
0.001
&292.6&0.64& &
412.8&0.91& &
604.8&1.33&
\\
0.01
&92.3&0.64& &
130.0&0.90& &
190.0&1.32&
\\
0.04
&46.0&0.64& &
64.7&0.90& &
94.3&1.31&
\\
0.1
&29.0&0.64& &
40.8&0.90& &
59.4&1.31&
\\
1.0
&9.0&0.63& &
12.8&0.90& &
19.4&1.30
\label{tab1pd}
\end{tabular}
\end{table}
\newpage
\begin{table}
\caption{
Cross sections $\sigma(\mbox{T}$-$\mbox{H}_\mu) =
\sigma_{\mbox{tr}}/10^{-20}$ cm$^2$ and rates 
$\lambda(\mbox{T}$-$\mbox{H}_\mu) =
\lambda_{\mbox{tr}}/10^{10}\mbox{s}^{-1}$
for $\mu$-transfer reaction
T \ + \ H$_{\mu} \rightarrow$ T$_{\mu}$ \ + \ H,
at different energies.}

\vspace{2mm}
\begin{tabular}{lccccccccc}
  \multicolumn{1}{l}{$E$(eV)}
& \multicolumn{1}{c}{$\sigma(\mbox{T}$-$\mbox{H}_\mu)$}
& \multicolumn{1}{c}{$\lambda(\mbox{T}$-$\mbox{H}_\mu)$}
& \multicolumn{1}{c}{ }
& \multicolumn{1}{c}{$\sigma(\mbox{T}$-$\mbox{H}_\mu)$}
& \multicolumn{1}{c}{$\lambda(\mbox{T}$-$\mbox{H}_\mu)$}
& \multicolumn{1}{c}{ }
& \multicolumn{1}{c}{$\sigma(\mbox{T}$-$\mbox{H}_\mu)$}
& \multicolumn{1}{c}{$\lambda(\mbox{T}$-$\mbox{H}_\mu)$}\\
  \multicolumn{1}{l}{ }
& \multicolumn{2}{c}{1s}
& \multicolumn{1}{c}{ }
& \multicolumn{2}{c}{1s+2s}
& \multicolumn{1}{c}{ }
& \multicolumn{2}{c}{1s+2s+2p}\\
\hline
0.001
&204.2&0.42& &
249.4&0.52& &
294.4&0.61&
\\
0.01
&64.3&0.42& &
78.5&0.51& &
92.6&0.60&
\\
0.04
&31.9&0.42& &
38.9&0.51& &
45.8&0.60&
\\
0.1
&19.9&0.41& &
24.3&0.50& &
28.6&0.60&
\\
1.0
&5.50&0.36& &
6.70&0.44& &
8.0&0.52
\label{tab2pt}
\end{tabular}
\end{table}
\vspace{2cm}
\begin{table}
\caption{
Cross sections $\sigma(\mbox{T}$-$\mbox{D}_\mu) =
\sigma_{\mbox{tr}}/10^{-20}$ cm$^2$ and rates 
$\lambda(\mbox{T}$-$\mbox{D}_\mu) =
\lambda_{\mbox{tr}}/10^{8}\mbox{s}^{-1}$
for $\mu$-transfer reaction 
T \ + \ D$_{\mu} \rightarrow$ T$_{\mu}$ \ + \ D,
at different energies.}

\vspace{2mm}
\begin{tabular}{lccccccccc}
  \multicolumn{1}{l}{$E$(eV)}
& \multicolumn{1}{c}{$\sigma(\mbox{T}$-$\mbox{D}_\mu)$}
& \multicolumn{1}{c}{$\lambda(\mbox{T}$-$\mbox{D}_\mu)$}
& \multicolumn{1}{c}{ }
& \multicolumn{1}{c}{$\sigma(\mbox{T}$-$\mbox{D}_\mu)$}
& \multicolumn{1}{c}{$\lambda(\mbox{T}$-$\mbox{D}_\mu)$}
& \multicolumn{1}{c}{ }
& \multicolumn{1}{c}{$\sigma(\mbox{T}$-$\mbox{D}_\mu)$}
& \multicolumn{1}{c}{$\lambda(\mbox{T}$-$\mbox{D}_\mu)$}\\
  \multicolumn{1}{l}{ }
& \multicolumn{2}{c}{1s}
& \multicolumn{1}{c}{ }
& \multicolumn{2}{c}{1s+2s}
& \multicolumn{1}{c}{ }
& \multicolumn{2}{c}{1s+2s+2p}\\
\hline
0.001
&4.58&0.77& &
5.05&0.84& &
13.7&2.3&
\\
0.01
&1.44&0.76& &
1.60&0.84& &
4.3&2.3&
\\
0.04
&0.71&0.75& &
0.78&0.83& &
2.14&2.26&
\\
0.1
&0.44&0.73& &
0.48&0.81& &
1.32&2.21&
\\
1.0
&0.1&0.44& &
0.1&0.5& &
0.3&1.5
\label{tab3dt}
\end{tabular}
\end{table}
\newpage
\begin{table}
\caption{Elastic cross sections for T-D$_\mu$ collision
in units of $10^{-20}$ cm$^2$ at different energies.}
\vspace{2mm}
\begin{tabular}{lcccc}
  \multicolumn{1}{l}{$E$(eV)}
& \multicolumn{1}{c}{Present results}
& \multicolumn{1}{c}{\cite{Fukuda-1990}}
& \multicolumn{1}{c}{\cite{Igarashi-1994}}
& \multicolumn{1}{c}{\cite{Lec-1997}}\\
  \multicolumn{1}{l}{ }
& \multicolumn{1}{c}{1s+2s+2p}
& \multicolumn{1}{c}{ }
& \multicolumn{1}{c}{ }
& \multicolumn{1}{c}{ }\\
\hline
0.001 & 1.2 & 1.7 & 1.63 & 2.014\\
0.01  & 1.3 & 2.3 & 2.15 & 3.605
\label{tab5el}
\end{tabular}
\end{table}
\vspace{4cm}
%
%
\begin{table}
\caption{Cross sections in unit $\pi a_0^2$
for positron transfer reaction
$\bar{\mbox{p}}$ \ + \ Ps $\rightarrow \bar{\mbox{H}} \ + \ e^-$;
$^{\dag} -$ the cross sections estimated from Figure 1
of Ref. [26].} 
\vspace{2mm}
\begin{tabular}{lcclcc}
  \multicolumn{1}{l}{$E$(eV)}
& \multicolumn{1}{c}{Present results}
& \multicolumn{1}{c}{\cite{Toshima-1994}}
& \multicolumn{1}{l}{$E(eV)$}
& \multicolumn{1}{c}{Present results}
& \multicolumn{1}{c}{\cite{Toshima-1994}}\\
  \multicolumn{1}{l}{ }
& \multicolumn{1}{c}{1s+2s+2p}
& \multicolumn{1}{c}{ }
& \multicolumn{1}{l}{ }
& \multicolumn{1}{c}{1s+2s+2p}
& \multicolumn{1}{c}{ }\\
\hline
0.1 & 1.5 & 2.3$^{\dag}$ & 1.0 & 3.2 & 3.5$^{\dag}$\\
0.5 & 2.0 & 2.6$^{\dag}$ & 2.0 & 1.7 & 3.7$^{\dag}$
\label{tab5ah}
\end{tabular}
\end{table}
%
%
%
%

\begin{thebibliography}{99}

\bibitem {Tresch2-1998}
S. Tresch, F. Mulhauser, C. Piller, L. A. Schaller, L. Schellenberg,
H. Schneuwly, Y. A. Thalmann, A. Werthmuller,
P. Ackerbauer, W. H. Breunlich, M. Cargnelli, B. Gartner, R. King,
B. Lauss, J. Marton, W. Prymas, J. Zmeskal, C. Petitjean,
M. Augsburger, D. Chatellard, J. P. Egger, E. Jeannet, T. von Egidy,
F. J. Hartmann, M. Muhlbauer, and W. Schott,
Phys. Rev. A {\bf 58}, 3528 (1998).

\bibitem {Tresch1-1998}
S. Tresch, R. Jacot-Guillarmod, F. Mulhauser, C. Piller, L. A. Schaller,
L. Schellenberg, H. Schneuwly, Y. A. Thalmann,
A. Werthmuller, P. Ackerbauer, W. H. Breunlich, M. Cargnelli, B. Gartner,
R. King, B. Lauss, J. Marton, W. Prymas,
J. Zmeskal, C. Petitjean, D. Chatellard, J. P. Egger, E. Jeannet,
F. J. Hartmann, and M. Muhlbauer, Phys. Rev. A {\bf 57}, 2496 (1998).

\bibitem {Thalmann-1998}
Y.-A. Thalmann, R. Jacot-Guillarmod, F. Mulhauser, L. A. Schaller,
L. Schellenberg, H. Schneuwly, S. Tresch, and A. Wertm\"uller,
Phys. Rev. A {\bf 57}, 1713 (1998).

\bibitem {Jacot-1998}
S. Tresch, R. Jacot-Guillarmod, F. Mulhauser, L. A. Schaller,
L. Schellenberg, H. Schneuwly, Y.-A. Thalmann, and
A. Werthm\"uller, Euro. Phys. J. D {\bf 2}, 93 (1998).

\bibitem {Mulhauser-1993}
F. Mulhauser and H. Schneuwly, J. Phys. B {\bf 26}, 4307 (1993);
L. Schellenberg, Hyperf. Interact. {\bf 82}, 513 (1993).

\bibitem {Breunlich-89}
W. H. Breunlich, P. Kammel, J.S. Cohen, and M. Leon,
Annu. Rev. Nucl. Part. Sci. {\bf 39}, 311 (1989).

\bibitem {Holz-1997}
M. H. Holzscheiter, G. Bendiscioli, A. Bertin, G. Bollen, M. Bruschi,
C. Cesar,
M. Charlton, M. Corradini, D. DePedis,
M. Doser, J. Eades, R. Fedele, X. Feng, F. Galluccio, T. Goldman,
J. S.  Hangst, R. Hayano, 
D. Horv\'ath, R. J. Hughes, N. S. P. King,
K. Kirsebom, H. Knudsen, V. Lagomarsino, R. Landua, G. Laricchia,
R. A. Lewis, E. LodiRizzini, M. Macri, G. Manuzio,
U. Marconi, M. R. Masullo, J. P. Merrison,
S. P. Moller, G. L. Morgan, M. M. Nieto,
M. Piccinini, R. Poggiani, A. Rotondi,
G. Rouleau, P. Salvini, N. Semprini, N. Cesari,
G. A. Smith, C. M. Surko,
G. Testera, G. Torelli, E. Uggerhoj, V. G. Vaccaro,
L. Venturelli, A. Vitale, E. Widmann, T. Yamazaki, Y. Yamazaki,
D. Zanello, and A. Zoccoli, Hyperfine Interact. {\bf 109}, 1 (1997).

\bibitem {Charlton-94}
J. Eades and F. J. Hartmann, Rev. Mod. Phys. {\bf 71}, 373 (1999);
M. Charlton, J. Eades, D. Horv\'ath, R. J. Hughes, and C. Zimmerman,
Phys. Rep. {\bf 241}, 65 (1994).

\bibitem {Hahn-1968}
Y. Hahn, Phys. Rev. {\bf 169}, 794 (1968).

\bibitem {Hahn-1972}
Y. Hahn and K. M. Watson, Phys. Rev. A {\bf 5}, 1718 (1972).

\bibitem {Faddeev-1961}
L. D. Faddeev, Zh. Eksp. Teor. Fiz. {\bf 39}, 1459 (1960)
[Sov. Phys.$-$JETP {\bf 12}, 1014 (1961)].

\bibitem {Sultanov-1999}
R. A. Sultanov, W. Sandhas, and V. B. Belyaev, Euro. Phys. J. D {\bf 5},
33 (1999).

\bibitem {Dzhel-1962}
V. P. Dzhelepov, P.F. Ermolov, E. A. Kushnirenko, V. I. Moskalev, and
S. S. Gershtein, Zh. Eksp. Theor. Fiz. {\bf 42}, 439 (1962)
[Sov. Phys.$-$JETP {\bf 15}, 306 (1962)].

\bibitem {Bleser-1963}
E. J. Bleser, E. W. Anderson, L. M. Lederman, S. L. Meyer,
J. L. Rosen, J. E. Rothberg, and I-T. Wang,
Phys. Rev. {\bf 132}, 2679 (1963).

\bibitem {Bertin-1972}
A. Bertin, M. Bruno, V. Vitale, A. Placci, and E. Zavattini,
Nuovo Cimento Lett. {\bf 4}, 449 (1972).

\bibitem {Mulhauser-1996}
F. Mulhauser, J. L. Beveridge, G. M. Marshall, J. M. Bailey,
G. A. Beer, P. E. Knowles, G. R. Mason, A. Olin, M. C.  Fujiwara,
T. M.  Huber, R. Jacot-Guillarmod,
P. Kammel, J. Zmeskal, S. K. Kim, A. R. Kunselman,
V. E. Markushin, C. J. Martoff, and C. Petitjean,
Phys. Rev. A {\bf 53}, 3069 (1996).

\bibitem {Bystr-1981}
V. M. Bystritsky, V. P. Dzhelepov, Z. V. Yershova, V. G. Zinov, 
V. K. Kapyshev, S. S. Mukhametgaleyeva, V. S. Nadezhdin, L. A. Rivkis,
A. I. Rudenko, V. I. Satarov, N. V. Sergeyeva, L. N. Somov,
V. A. Stolupin, and V. V. Filchenkov,
Zh. Eksp.Teor. Fiz. {\bf 80}, 1700 (1980)
[Sov. Phys.$-$JETP {\bf 53}, 877 (1981)].

\bibitem {Jones-83}
S. E. Jones, A. N. Anderson, A. J. Caffrey, J. B. Walter, K. D. Watts,
J. N. Bradbury, M. Leon, H. R. Maltrud, and M. A. Paciotti,
Phys. Rev. Lett. {\bf 51}, 1757 (1983).

\bibitem {Breun-87}
W. H. Breunlich, M. Cargnelli,
P. Kammel, J, Marton, N. Naegele, P. Pawlek,
A. Scrinzi,
J. Werner, J. Zmeskal, J. Bistirlich, K. M. Crowe, M. Justice,
J. Kurck, C. Petitjean, R. H. Sherman, H. Bossy, H. Daniel,
F. J. Hartmann, W. Neumann, and G. Schmidt,
Phys. Rev. Lett. {\bf 58}, 329 (1987).

\bibitem {Cohen-1991}
J. S. Cohen and M. C. Struensee, Phys. Rev. A {\bf 43}, 3460 (1991).

\bibitem {Adam-1992}
A. Adamczak, C. Chiccoli, V. I. Korobov, V. S. Melezhik, P. Pasini,
L. I. Ponomarev, and J. Wozniak, Phys. Lett. B {\bf 285}, 319 (1992).

\bibitem {Fukuda-1990}
H. Fukuda, T. Ishihara, and S. Hara, Phys. Rev. A {\bf 41}, 145 (1990).

\bibitem {Kamimura-1993}
Y. Kino and M. Kamimura, Hyperfine Interact. {\bf 82}, 45 (1993).

\bibitem {Igarashi-1994}
A. Igarashi, N. Toshima, and T. Shirai, Phys. Rev. A {\bf 50}, 4951 (1994).

\bibitem {Lec-1997}
A. Boukour, R. N. Hewitt, and Ch. Leclercq-Willain,
J. Phys. B {\bf 29}, 4309 (1996).

\bibitem {Toshima-1994}
A. Igarashi, N. Toshima, and T. Shirai, J. Phys. B {\bf 27}, L497 (1994).

\bibitem {Mitroy-1997}
J. Mitroy and G. Ryzhikh, J. Phys. B {\bf 30}, L371 (1997).

\bibitem {Mott-1965}
N. F. Mott and H. S. W. Massey, {\it The theory of
atomic collisions} (Clarendon, London, 1965).

\bibitem {Lauss-96}
B. Lauss, P. Ackerbauer, W. H. Breunlich, B. Gartner, M. Jeitler,
P. Kammel, J. Marton, W. Prymas, J. Zmeskal, D. Chatellard,
J. P. Egger, E. Jeannet, H. Daniel, A. Kosak, F. J. Hartmann,
and C. Petitjean, Phys. Rev. Lett. {\bf 76}, 4693 (1996).

\bibitem {Krav-1998}
A. V. Kravtsov and A. I. Mikhailov, Phys. Rev. A {\bf 58}, 4426 (1998).

\end{thebibliography}
\end{document}